\documentclass[a4paper]{jpconf}
\usepackage{graphicx}
\usepackage{iopams}
\begin{document}
\title{High Field determination of superconducting fluctuations in high-T$_c$ cuprates}

\author{F. Rullier-Albenque$^1$, H. Alloul$^2$, D. Colson$^1$ and A. Forget $^1$}

\address{$^1$ Service de Physique de l'Etat Condens\'e, Orme des Merisiers, CEA Saclay (CNRS URA 2464), 91191 Gif sur Yvette cedex, France}
\address{$^2$ Laboratoire de Physique des Solides, UMR CNRS 8502, Universit\'e Paris Sud, 91405 Orsay, France}

\ead{florence.albenque-rullier@cea.fr}

\begin{abstract}
Large pulsed magnetic fields up to 60 Tesla are used to suppress the contribution of superconducting
fluctuations (SCF) to the ab-plane conductivity above $T_c$ in a series of YBa$_2$Cu$_3$O$_{6+x}$ 
single crystals. 
The fluctuation conductivity is found to vanish nearly exponentially with temperature, allowing us to 
determine precisely the field $H_c^{\prime}(T)$ and the temperature $T_c^{\prime}$ above which the SCFs 
are fully suppressed. $T_c^{\prime}$ is always found much smaller than the pseudogap temperature 
A careful investigation near optimal doping shows that $T_c^{\prime}$ is higher than the 
pseudogap $T^{\star}$, which indicates that pseudogap cannot be assigned to preformed pairs.
For nearly optimally doped samples, the fluctuation conductivity can be accounted for by gaussian
fluctuations following the Ginzburg-Landau scheme. A phase fluctuation contribution might be invoked for the 
most underdoped samples in a $T$ range which increases when controlled disorder is introduced by 
electron irradiation. Quantitative analysis of the fluctuating magnetoconductance allows us to determine 
the critical field $H_{c2}(0)$ which is found to be quite similar to $H_c^{\prime}(0)$ and to 
increase with hole doping.
Studies of the incidence of disorder on both $T_c^{\prime}$ and $T^{\star}$ enable us to propose a 
three dimensional phase diagram including a disorder axis, which allows to explain most observations 
done in other cuprate families.
\end{abstract}

\section{Introduction}
\begin{figure}[h]
\includegraphics[width=18pc]{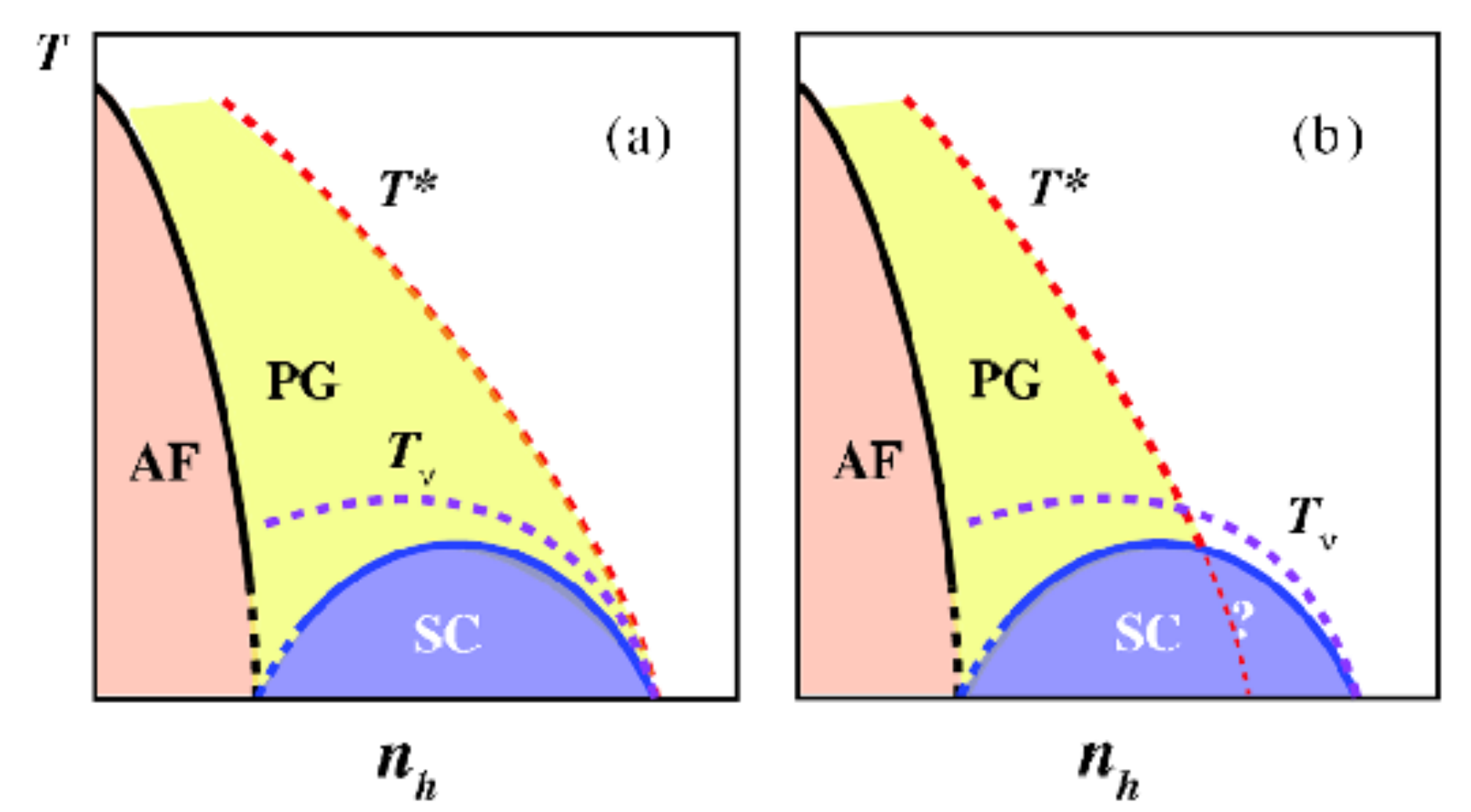}\hspace{2pc}%
\begin{minipage}[b]{16pc}\caption{\label{fig:phase-diagram}Different scenarios for the phase diagram of 
the high-$T_c$ cuprates. While in (a) $T^*$ merges with $T_c$ in the strongly overdoped regime, 
in (b) $T^{\star}$ intersects the superconducting dome 
near optimal doping. $T_{\nu}$ represents the onset of the Nernst signal. From ref.\cite{Alloul-EPL2010}}
\end{minipage}
\end{figure}

One of the most puzzling feature of the high-$T_c$ cuprates is the existence of the so-called pseudogap phase in the underdoped region of their phase diagram. After the first evidence of an anomalous drop of the spin susceptibility detected by NMR experiments in underdoped YBCO well above $T_c$ \cite{Alloul-1989}, a lot of unusual properties have been observed in the pseudogap phase \cite{Timusk}. However its exact relationship with  superconductivity is still highly debated. In fact there does not exist up to now a unique representation for the pseudogap line $T^{\star}$ as illustrated in Fig.\ref{fig:phase-diagram}. Either $T^*$ is found to merge with the superconducting dome in the overdoped part of the phase diagram, or to cross it near optimal doping. In the first case, it has been proposed that the pseudogap could be ascribed to the formation of superconducting pairs with strong phase fluctuations \cite{Emery}. This scenario has been supported by the observation of a large Nernst effect and of diamagnetism above $T_c$, which delineates another line $T_{\nu}$ below which strong superconducting fluctuations and/or vortices persist in the normal state \cite{Wang-PRB2006}. In the second approach, the pseudogap and the superconducting phases arise from different, even competing, underlying mechanisms and are associated with different energy scales \cite{Hufner}.

\textit{While superconducting fluctuations (SCF) are expected to be especially large in these anisotropic compounds with low superfluid density}, there is not a clear consensus about the temperature range above $T_c$ in which SCFs survive. In this paper, we will present our results on superconducting fluctuations for a series of
YBa$_2$Cu$_3$O$_{6+x}$ single crystals from underdoping to slight overdoping. We have used an original
method based on the measurements of the magnetoresistance in high pulsed magnetic fields. We have initiated this method in ref.\cite{FRA-HF} in underdoped compounds, and have done complementary measurements for various O contents \cite{FRA-PRB2011}. In this later paper we gave an extensive report on the data and their analysis. Here we present a simplified comprehensive summary of our main results and discuss them in the context of recent experimental results reported by others at this conference. 

The principle of our study is to use high magnetic fields to determine the normal state resistivity and to 
extract with high accuracy the superconducting
fluctuations (SCF) contributions to the conductivity and their dependences as a function of temperature 
and magnetic field (section 2). We are thus able to determine the threshold values of the magnetic field 
$H_{c}^{\prime}$ and temperature $T_{c}^{\prime}$ above which the normal state is completely restored. 
In the same set of transport data, we could compare the values of $T_{c}^{\prime}$ and of the 
pseudogap temperature $T^{\star}$ as a function of doping \cite{Alloul-EPL2010} (sec.3). We have added in the present paper another comparison of these two temperature scales by using previous NMR data to determine $T^{\star}$ \cite{Alloul-1989}. We will also show how 
our results can be analysed in the framework of the Ginzburg-Landau approach, making it possible to 
extract microscopic parameters of the superconducting state such as the zero-temperature coherence 
length (sec.4-5). 
The effect of disorder introduced by electron irradiation at low temperature will be also presented (sec.6).

\section{Experimental}
Details on the experimental conditions concerning the different single crystals and the high-field
experiments as well as the method used to extract the SCF contribution to the conductivity are given 
in ref.\cite{FRA-PRB2011}. Four different single crystals of YBa$_2$Cu$_3$O$_{6+x}$ have been studied. 
They are labelled with respect to their critical temperatures measured at the mid-point of the resistive
transition: two underdoped samples UD57 and UD85, an optimally doped sample OPT93.6 and a slightly 
overdoped one OD92.5, corresponding to oxygen contents of approximately 6.54, 6.8, 6.91 and 6.95
respectively. Some of these samples have been irradiated by electrons at low $T$, which allows us 
to introduce a well controlled concentration of defects in the CuO$_2$ planes \cite{Legris}.

\begin{figure}[h]
\includegraphics[width=20pc]{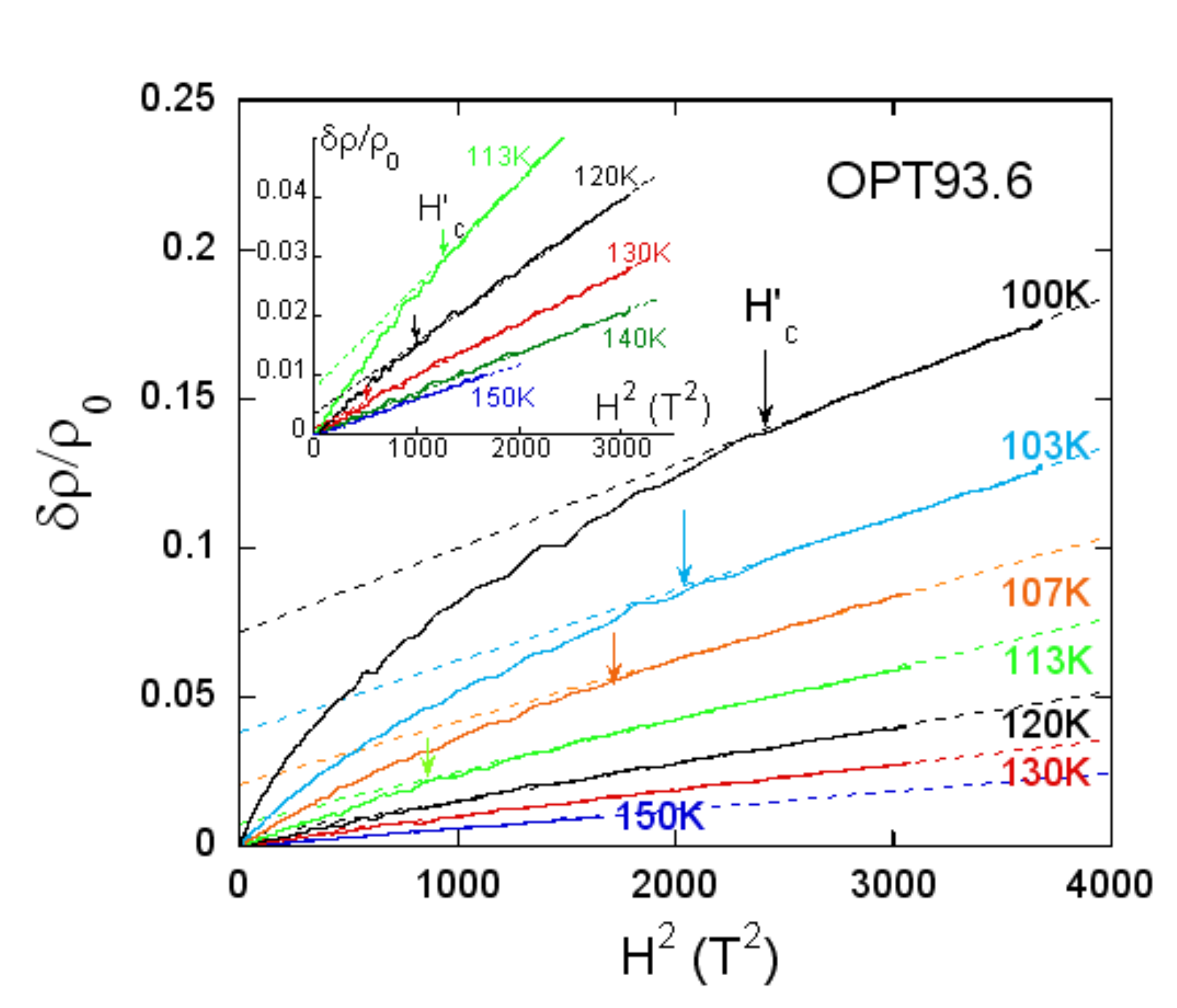}
\begin{minipage}[b]{18pc}\caption{\label{fig:MR-OPT}Field variation of the resistivity increase normalized to its zero-field value $\delta\rho/\rho_0$ plotted versus $H^{2}$ for decreasing temperatures down to $T \backsimeq T_c$ in the optimally doped sample OPT93.6. The inset shows an enlargement of the curves for the highest temperatures. (from ref.\cite{FRA-PRB2011}).}
\end{minipage}
\end{figure}

The transverse MR of the different samples have been measured in a pulse field magnet up to 60T at 
the LNCMI in Toulouse. An example of the transverse MR curves measured on the OPT93.6 sample is illustrated
in fig.\ref{fig:MR-OPT} for $T$ ranging from above $T_c$ to 150K. In the normal state well above $T_c$, it 
was shown that the transverse MR increases as $H^2$ both in optimally doped and in underdoped YBCO \cite{Harris}. This is indeed what is found also here for $H$ up 
to 60T and for $T\gtrsim140$K (see inset of fig.\ref{fig:MR-OPT}). At lower $T$, some downward departure 
from this $H^2$ behavior is observed for low values of $H$ which we attribute to the destruction of SCFs by 
the magnetic field. The normal state behavior is only restored above a threshold field $H_{c}^{\prime}(T)$
which increases with decreasing temperatures. In ref.\cite{FRA-PRB2011}, we have shown that the magnetoresistance coefficients measured at low $T$ above $H_{c}^{\prime}(T)$ are in continuity with those measured at higher temperatures and low field, which is a strong indication that the effect of the field is to restore the normal state.

This experimental approach allows us to single out the 
normal state properties and determine the SCF contributions to the transport. In particular, the
extrapolation down to $H=0$ of the $H^2$ normal state MR above $H_{c}^{\prime}(T)$ gives us the value of 
the normal state resistivity $\rho_{n}(T)$. The way to extract the fluctuating conductivity and its
dependence with temperature $\Delta\sigma_{SF}(T,0)$ and magnetic field $\Delta\sigma_{SF}(T,H)$ is 
explained in details in ref.\cite{FRA-PRB2011}.

\section{Onset of superconducting fluctuations and pseudogap}
\label{sec:onset and PG}
\begin{figure}[h]
\begin{minipage}{18pc}
\includegraphics[width=18pc]{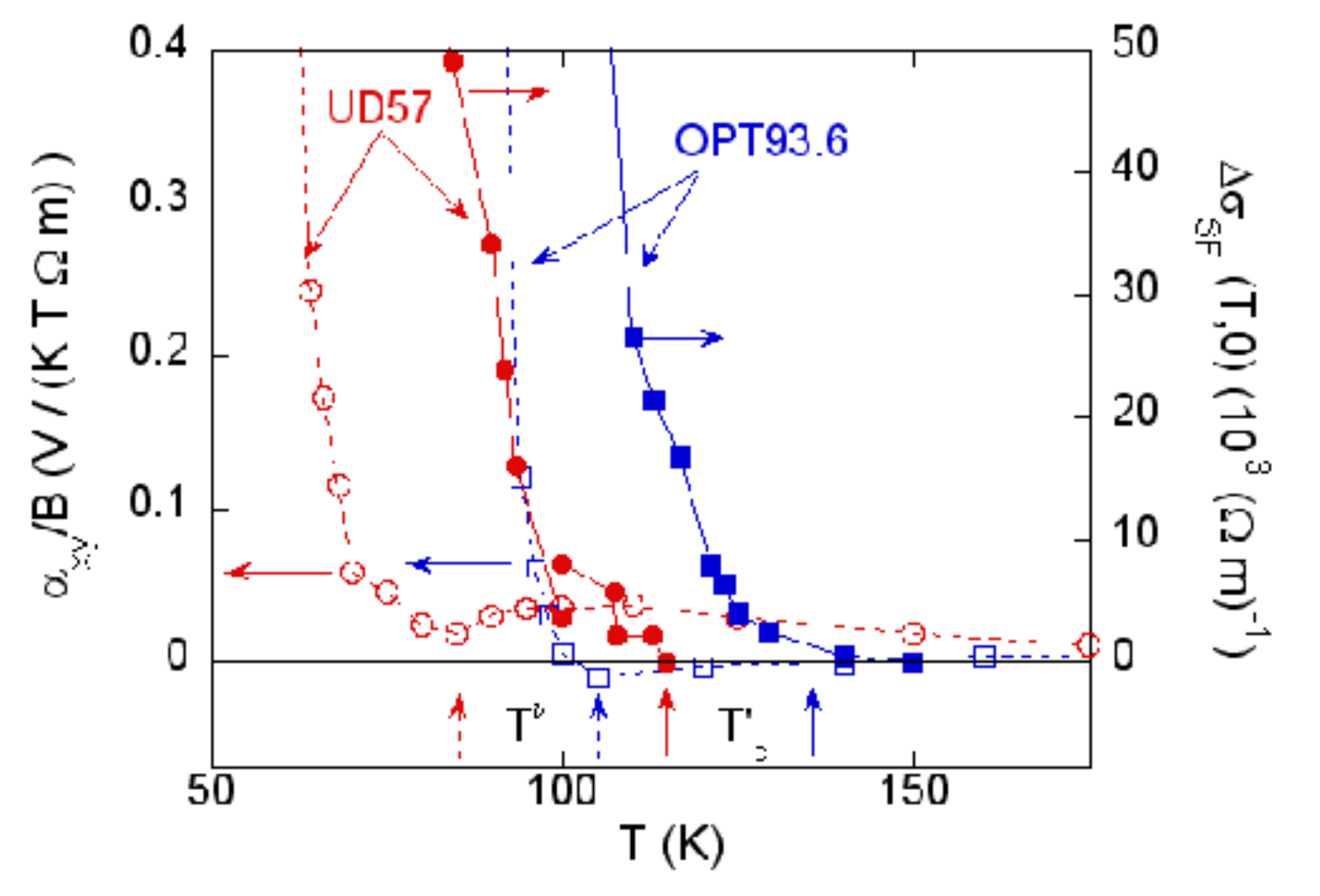}
\caption{\label{fig:Nernst-dsigma}Comparison of the onset temperature for SCFs extracted from Nernst 
measurements (open symbols) and from $\Delta\sigma_{SF}(T,0)$ (full symbols)
for the YBCO OPT93.6 (squares) and UD57 (circles) samples (from ref.\cite{FRA-PRB2011}). Lines are guides for the eyes.}
\end{minipage}\hspace{4pc}%
\begin{minipage}{18pc}
\includegraphics[width=18pc]{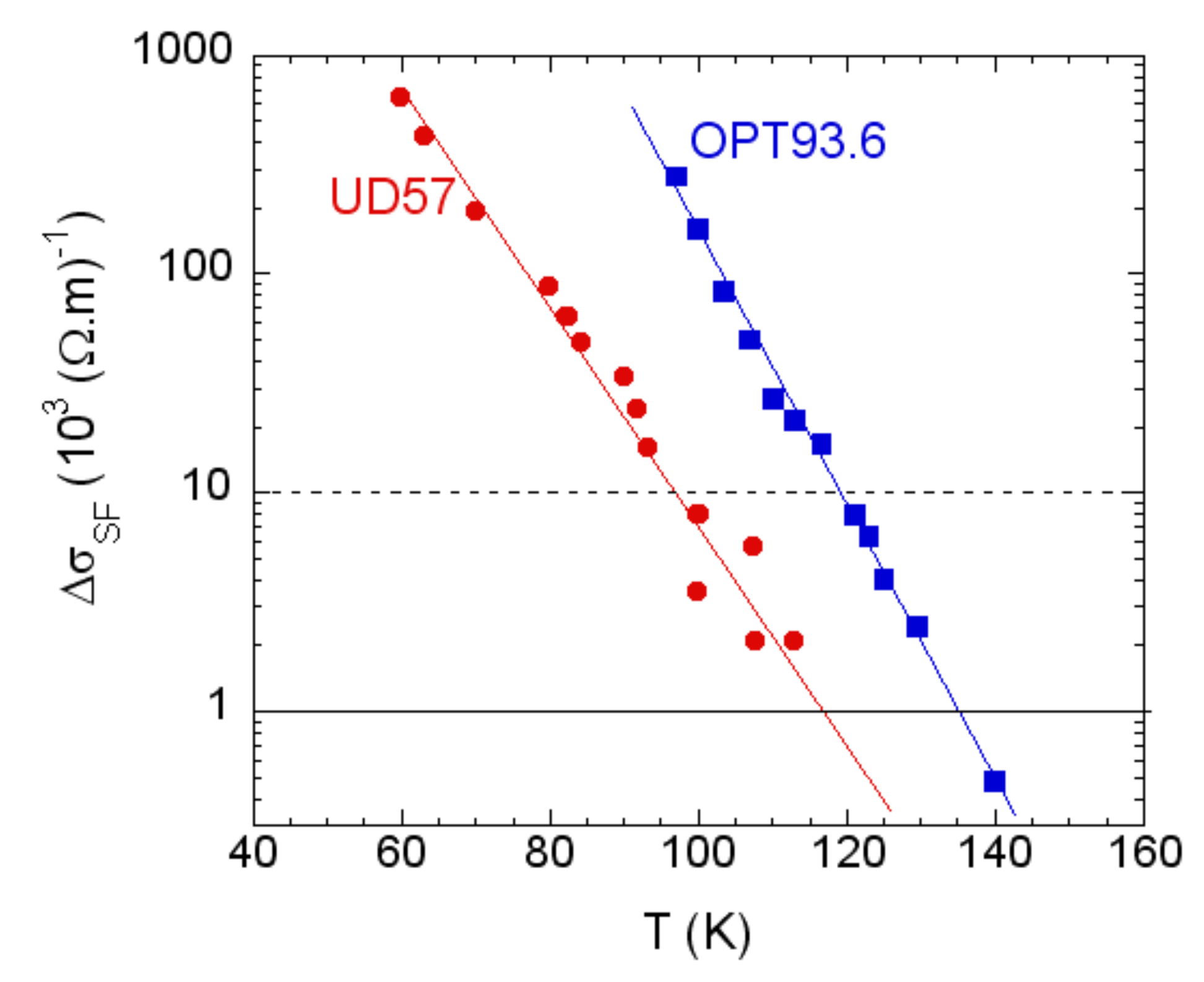}
\caption{\label{fig:exp-decay}$\Delta\sigma_{SF}(T,0)$ for the OPT93.5 and UD57 samples in a semi-logarithmic
scale. $T_{c}^{\prime}$ is defined as the temperature above 
which $\Delta\sigma_{SF}(T,0)$ is lower than 1 x $10^{3}(\Omega m)^{-1}$ (full line). For a lower sensitivity
($\Delta\sigma_{SF}(T,0)=10$ x $10^{3}(\Omega m)^{-1}$- dashed line) the values of $T_{c}^{\prime}$ 
would be smaller by 15-20K.}
\end{minipage} 
\end{figure}

The $T$ dependences of the zero-field superconducting conductivities $\Delta\sigma_{SF}(T,0)$ are reported 
in Fig.\ref{fig:Nernst-dsigma} for the OPT93.6 and UD57 samples and compared to the off-diagonal Peltier conductivity deduced from the Nernst measurements on the same samples \cite{FRA-Nernst}. 
We observe that $T_{c}^{\prime}$ is always found larger that the onset of Nernst signal $T_{\nu}$. 
This might come from the difficulty to choose a good criterion to determine $T_{\nu}$ as the minimum 
in $\alpha_xy/B$ may hide the real onset of SCFs. On the contrary, in Fig.\ref{fig:exp-decay} 
where $\Delta\sigma_{SF}(T,0)$ are plotted in a semi-logarithmic scale, one can note that this quantity 
vanishes very fast. This allows us to define a precise criterion to determine the onset temperature 
$T_{c}^{\prime}$, corresponding here to  $\Delta\sigma_{SF}(T,0)$=1 x $10^{3}(\Omega m)^{-1}$. 
Of course decreasing or improving the sensitivity for SCF detection will result in decrease or increase 
of $T_{c}^{\prime}$, 
which might explain the different temperature ranges of SCFs derived from different experimental probes. 
If we were able to improve our sensitivity by an order of magnitude, the values of $T_c^{\prime}$ could only be increased by $\sim15$K  and would correspond to an extremely low SCF
contributions to the conductivity, about four orders of magnitude lower than at measured at $T_c$. 

In the case of the optimally doped compound, our value of $T_c^{\prime}$ is in very good agreement with the onset temperature determined by magnetic susceptibility \cite{Li}. However these values are found larger than those determined by microwave ac conductivity measurements \cite{Grbic}. This can be likely ascribed to the fact that a field of 16T is assumed to be sufficient to suppress all superconductivity above the zero-field $T_c$ in this latter work, which is clearly in contradiction with our results. On the other hand, recent terahertz conductivity measurements in LSCO show that the signatures of the fluctuations only persist in a very narrow range, at most 16K above $T_c$ \cite{Bilbro}. This has to be contrasted with determinations from Nernst measurements \cite{Wang-PRB2006} or high field magnetoresistance measurements comparable to ours \cite{Rourke} which give much larger onset temperatures. This clearly indicates that these different types of experiments do not detect the SCFs with a similar sensitivity as ours, or that they probe different types of SCFs (namely phase versus amplitude). 

\begin{figure}[h]
\begin{minipage}{18pc}
\includegraphics[width=18pc]{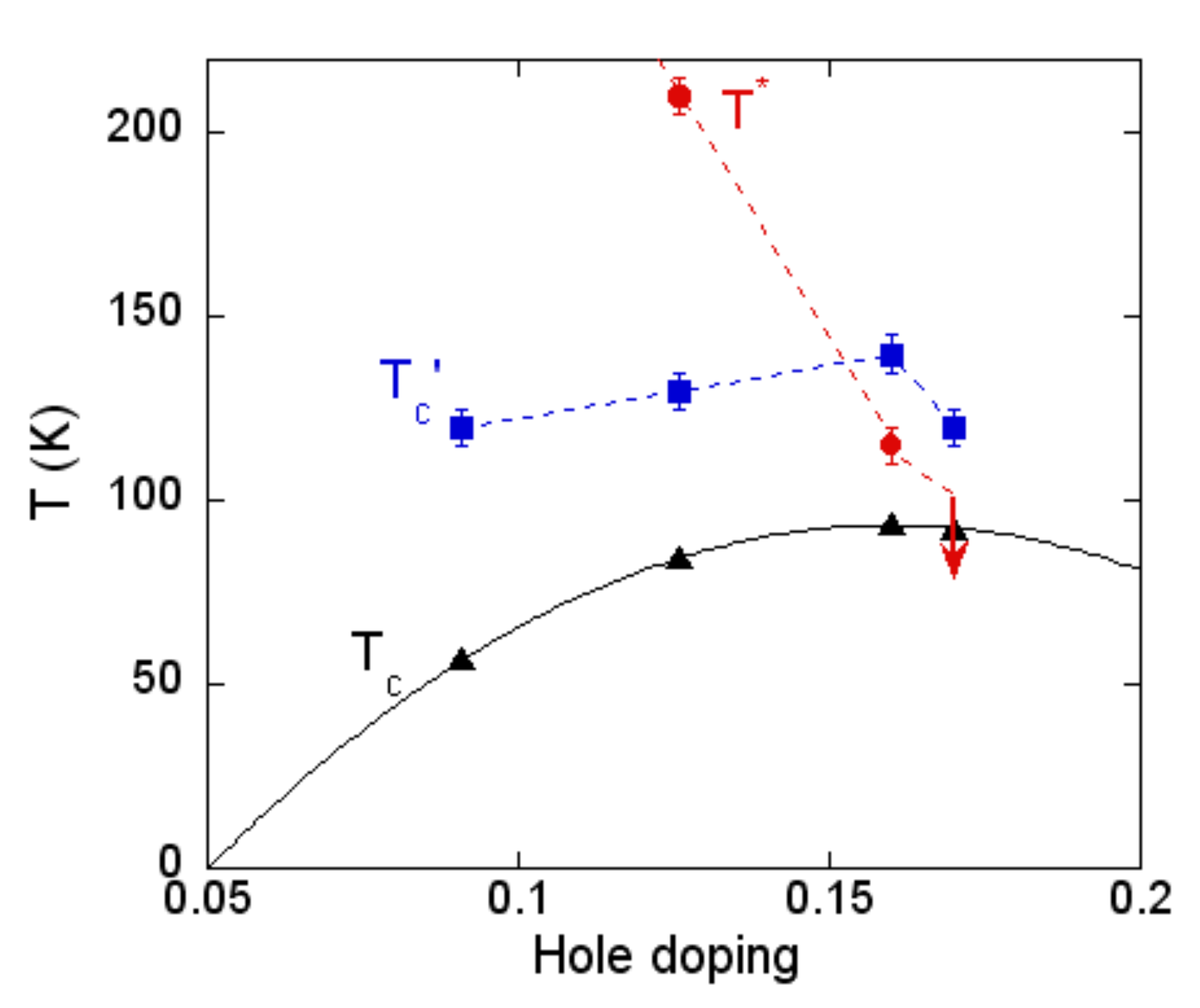}
\caption{\label{fig:onset-PG}The values of $T_{c}^{\prime}$ (\fullsquare) and $T^{\star}$ (\fullcircle) are plotted versus the hole doping for the four samples studied (from ref.\cite{FRA-PRB2011}). The solid line indicates the superconducting dome. Contrary to $T_{c}^{\prime}$ that is rather insensitive to hole doping, $T^{\star}$ is found to decrease with increasing doping and intersects the $T_{c}^{\prime}$ line near optimal doping \cite{Alloul-EPL2010}.}
\end{minipage}\hspace{4pc}%
\begin{minipage}{18pc}
\includegraphics[width=18pc]{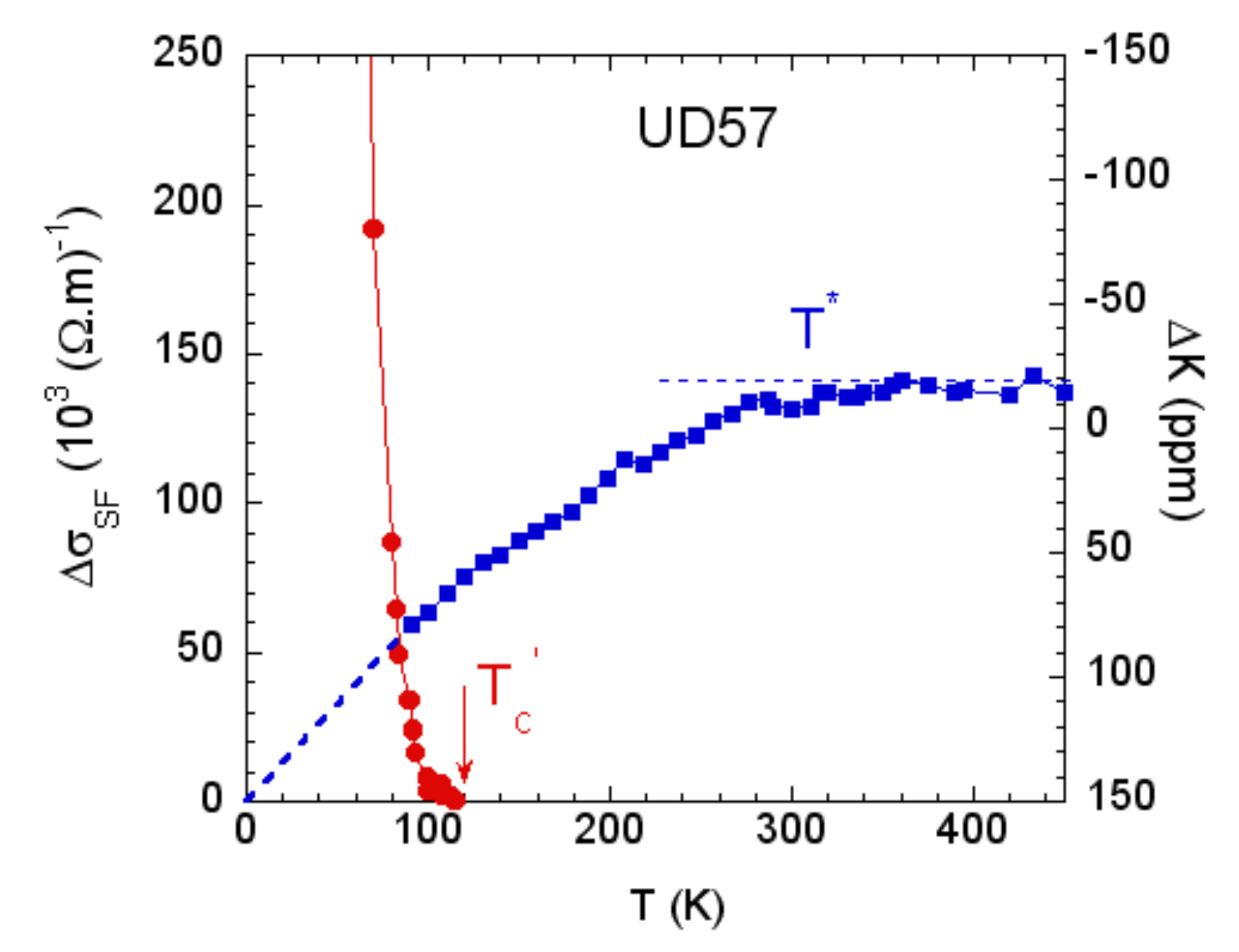}
\caption{\label{fig:RMN-SCF}The fluctuation conductivity (\fullcircle, left scale) is compared to the Y NMR Knight shift \cite{Alloul-1989}, \cite{Alloul-1993} (\fullsquare, right scale) for the UD57 sample. It is remarkable to see that nearly half of the susceptibility has already been lost at the onset of SCFs.}
\end{minipage} 
\end{figure}

Independently of these remarks, all the recent experimental data now point \textit{to an onset temperature for superconducting fluctuations well below the pseudogap temperature}.
One can also observe in Fig.\ref{fig:onset-PG} that $T_{c}^{\prime}$ is only slightly dependent on 
hole doping, increasing from $\backsim120$K to $\backsim140$K from the UD57 sample to the optimally 
doped one OPT93.6. This is very similar to what has been found from Nernst or magnetization experiments 
in Bi2212 \cite{Wang-PRL2}. However this strongly contrasts with the pseudogap temperature $T^{\star}$ which decreases with increasing doping. 

In order to compare the extension of the SCFs with respect to the opening of the pseudogap, the drop of the 
electronic susceptibility as measured by the Y NMR Knight shift \cite{Alloul-1993} is reported in 
Fig.\ref{fig:RMN-SCF} together with the fluctuation conductivity for the UD57 sample. These data clearly evidence 
that the electronic states lost by the opening of the pseudogap at $T^{\star}$ are not redistributed into the 
formation of preformed pairs in a precursor superconducting state as nearly half of the susceptibility has already 
been suppressed at $T_c^{\prime}$. This strongly indicates that these two states are not related. The situation 
is more delicate for the optimally doped sample for which 
$T^{\star}$ becomes comparable to $T_{c}^{\prime}$. In this case, a careful examination of the resistivity
data allowed us to determine both the onset of SCFs and the pseudogap 
temperature with the same experimental sensitivity criterion \cite{Alloul-EPL2010}. 
The fact that the $T_{c}^{\prime}$ line crosses the pseudogap line near optimal doping, as reported in 
Fig.\ref{fig:onset-PG}, unambiguously proves that the pseudogap phase cannot be a precursor state 
for superconductivity.

\section{Quantitative analysis of the paraconductivity in the Ginzburg-Landau approach}
\label{sec:Quant_analysis}
\begin{figure}[h]
\includegraphics[width=16pc]{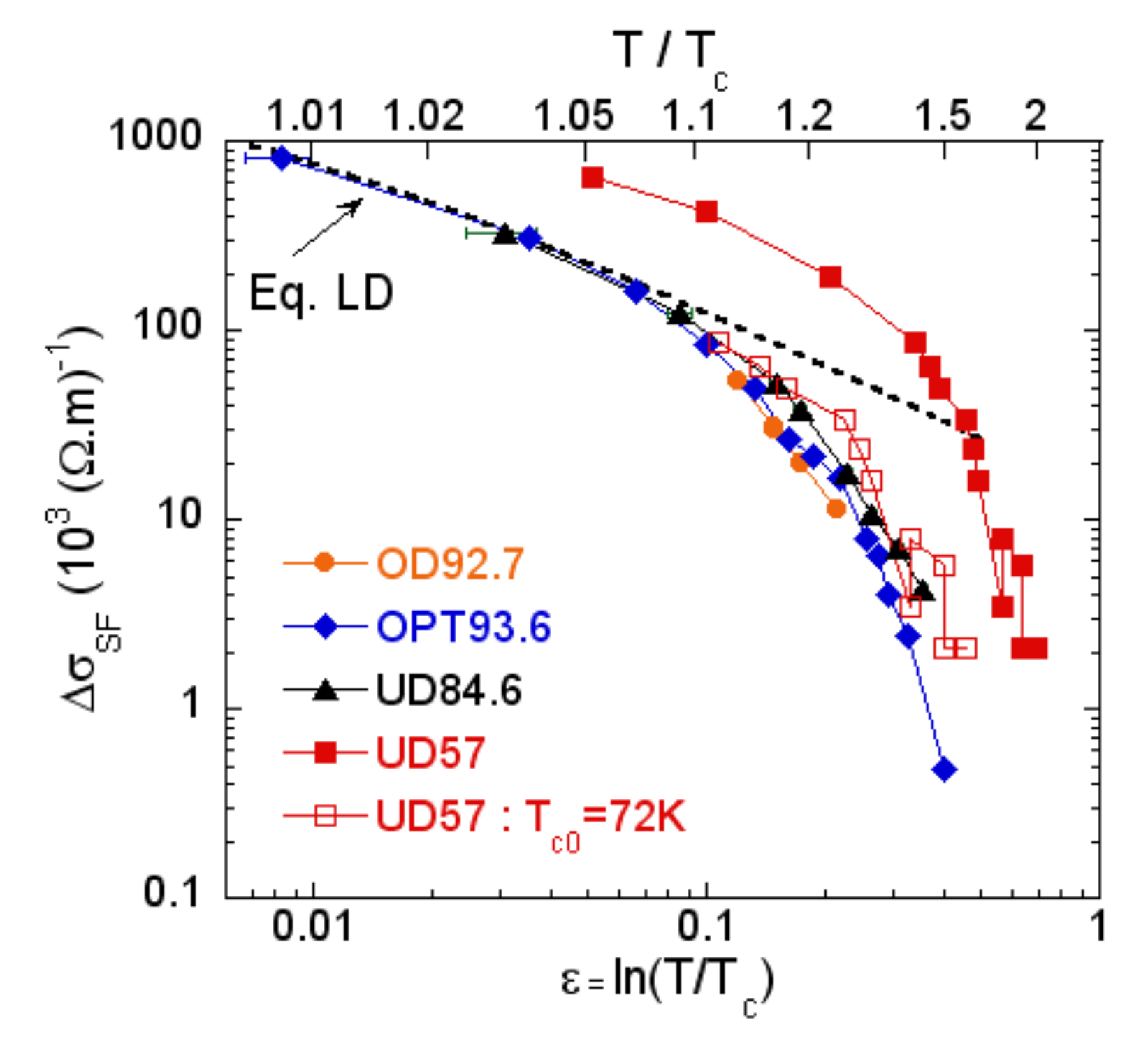}\hspace{2pc}%
\begin{minipage}[b]{20pc}\caption{\label{Fig:LD}Superconducting fluctuation conductivity $\Delta\sigma_{SF}$ for the four pure samples 
studied plotted versus $\epsilon=ln(T/T_{c})$ (from ref.\cite{FRA-PRB2011}). Values of $T_{c}$ have been taken 
here at the midpoint of the resistive transition, and error bars for $\epsilon$ using the onset and offset values 
of $T_{c}$ are indicated. The dashed line represents the expression of Eq.(\ref{Eq.LD})
with $s=11.7$\AA , and $\xi_{c}(0) \simeq 0.9$\AA . The data for the most underdoped sample can be matched with the other ones if one takes $T_{c0}=72$K for the actual $T_c$ instead of 57.1K. Full lines are guides to the eye.}
\end{minipage}
\end{figure}

The variations of $\Delta\sigma_{SF}(T)$ are reported versus $\epsilon=\ln(T/T_c)$ in fig.\ref{Fig:LD} 
for the four hole  dopings studied. Except for the UD57 sample, it is striking to see that the 
experimental data collapse on a single curve. For $\epsilon\lesssim 0.1$ these results can be well 
accounted for by gaussian fluctuations within the Ginzburg-Landau (GL) theory \cite{Larkin-Varlamov}. 
In this approach the excess fluctuating 
conductivity, called here paraconductivity, is related to the temperature dependence of $\xi(T)$, 
the superconducting correlation length of the short-lived Cooper pairs, which is expected to diverge 
with decreasing temperature as:
\begin{equation}
\xi(T)=\xi(0)/\sqrt{\epsilon},
\label{xi(T)} 
\end{equation}
where $\xi(0)$ is the zero-temperature coherence length and $\epsilon=\ln (T/T_c) \backsimeq (T-T_c)/T_c$ 
for $T \gtrsim T_c$. More generally, the temperature dependence of $\Delta\sigma_{SF}(T)$ is given by 
the Lawrence-Doniach (LD) expression which takes into account the layered structure of the high-$T_c$
cuprates \cite{Lawrence-Doniach}:
\begin{equation}
\Delta\sigma^{LD}(T)=\frac{e^{2}}{16\hslash s} \frac{1}{\epsilon \sqrt{1+2\alpha}},
\label{Eq.LD}
\end{equation}
where the coupling parameter $\alpha = 2 (\xi_{c}(T)/s)^{2}$ with $\xi_{c}(T)=\xi_{c}(0)/ \sqrt{\epsilon}$.
Sufficiently far from $T_{c}$, one expects $\xi_{c}(T) \ll s$ and Eq.\ref{Eq.LD} reduces to the 
well-known 2D Aslamazov-Larkin expression: 
\begin{equation}
\Delta\sigma^{AL}(T)=\frac{e^{2}}{16\hslash s} \epsilon^{-1}=\frac{e^{2}}{16\hslash s}\frac{\xi^{2}(T)}{\xi^{2}(0)}.
\label{Eq.AL}
\end{equation}
The only parameters in this expression are the value of the interlayer distance $s$ and the 
value taken for $T_c$ which can have a huge incidence on the shape of the curve especially 
for $(T-T_{c})/T_{c}<0.01$. It can be seen in Fig.\ref{Fig:LD} that the data are reasonably fitted by the LD expression (Eq.\ref{Eq.LD}) in the small 
temperature range $0.03 \leq \epsilon \leq 0.1$ if one takes $\xi_{c}(0) \simeq 0.9$\AA . We have 
assumed here, as usually done, that the CuO$_{2}$ bilayer constitutes the basic two-dimensional unit, 
and $s$ is then taken as the unit-cell size in the $c$ direction: $s=11.7$\AA .

For the UD57 sample, $\Delta\sigma_{SF}(T)$  is found  to be much 
larger (about a factor four at $\epsilon=0.05$) than for the other doping contents. 
Quite surprisingly, it is possible to recover a good matching with these
latter data by assuming a effective mean field temperature $T_{c0}$ different from the actual $T_c$. 
This is illustrated by the empty symbols in fig.\ref{Fig:LD} using $T_{c0}=72$K. This points to an 
additional origin of SCFs below $T_{c0}$ which might be ascribed to phase fluctuations of the order parameter. This is discussed in more details in ref.\cite{FRA-PRB2011}.

For all the samples, one can see in fig.\ref{Fig:LD} that $\Delta\sigma_{SF}(T)$ vanishes very rapidly 
for $\epsilon \gtrsim 0.1$. This behaviour which has been noticed previously in many studies is 
particularly well defined here given the method used to extract the fluctuating conductivity. 
An extension of the AL theory including short wave length fluctuations \cite{Reggiani} has been invoked 
to explain the steeper decrease of $\Delta\sigma_{SF}(T)$. This has also been treated by inserting a 
cutoff phenomenologically, implying that the density of fluctuating pairs vanishes at $T_{c}^{\prime}$ as detailed in ref.\cite{FRA-PRB2011}.

\section{Field variation of the SCF conductivity: $H_{c}^{\prime}$ and upper critical 
field $H_{c2}$}
\label{sec:Fields}
From the data reported in Fig.\ref{fig:MR-OPT}, we can also extract the magnetic field $H_{c}^{^\prime}(T)$ 
above which the MR recovers a $H^2$ dependence, which we take as the sign that the normal state is 
completely restored. As $T$ decreases, it becomes 
difficult to ascertain that the normal state is fully reached when $H_{c}^{^\prime}(T)$ becomes 
comparable to the highest available field. This makes it difficult to precisely deduce values 
of $H_{c}^{^\prime}(T)$ larger than 45T.

\begin{figure}[h]
\begin{minipage}{18pc}
\includegraphics[width=16pc]{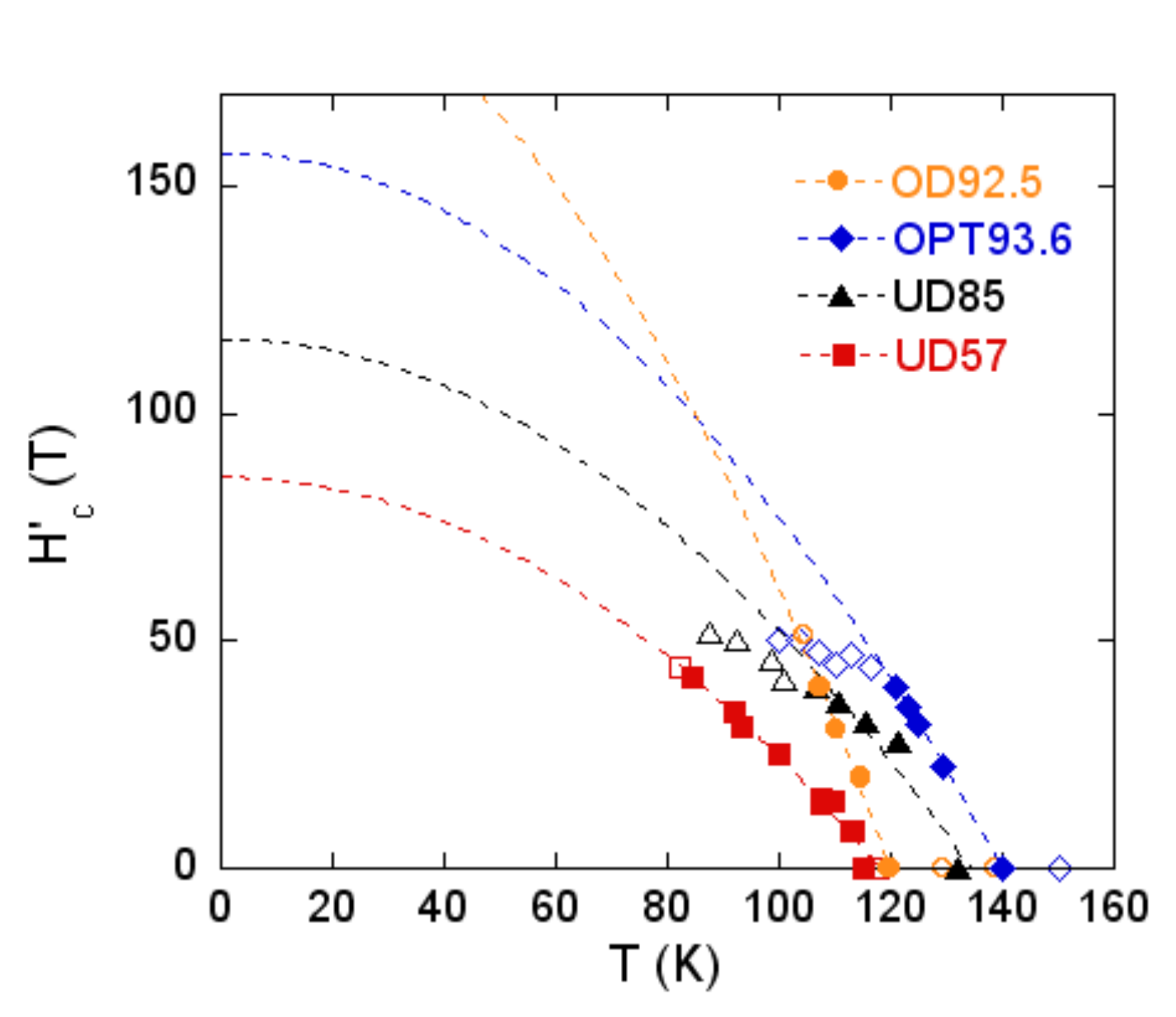}
\caption{\label{Fig:Hcprime}The field $H_{c}^{\prime}$, at which the SC
fluctuations disappear and the normal state is fully restored, is plotted versus 
$T$ for the four pure samples studied. Dashed lines represent the fitting curves 
to Eq.\ref{Eq.H'c-T} using data with closed symbols. 
When $H_{c}^{\prime}(T)\gtrsim 40T$ (empty symbols), the data are somewhat underestimated  as the  maximum 
applied field is not sufficient to restore the normal state.(from ref.\cite{FRA-PRB2011}).}
\end{minipage}\hspace{2pc}%
\begin{minipage}{18pc}
\includegraphics[width=18pc]{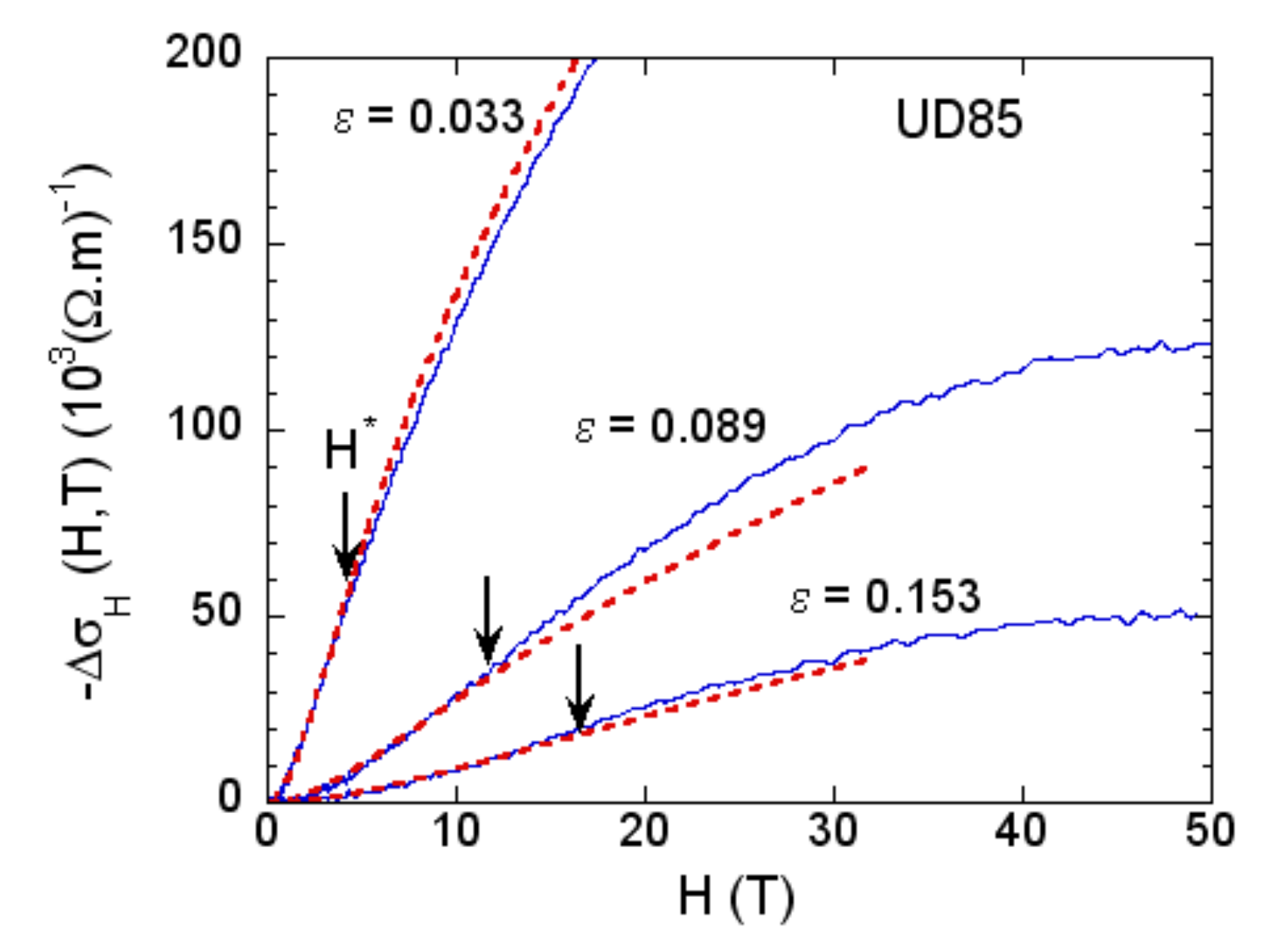}
\caption{\label{Fig:fit-ALO} Evolution of the fluctuation magnetoconductivity 
$-\Delta\sigma_{H}(T,H) =\Delta\sigma_{SF}(T,0)-\Delta\sigma_{SF}(T,H)$
as a function of $H$ for the UD85 sample at different temperatures: 87.5, 92.4 and 98.6K. 
The dotted lines represent the computed results from the ALO contribution with  $H_{c2}(0)=125(5)$T. 
They deviate from the data beyond the $H^{\star}$ field values shown by arrows \cite{FRA-PRB2011}.}
\end{minipage} 
\end{figure}

The evolution of $H_{c}^{^\prime}(T)$ are plotted in fig.\ref{Fig:Hcprime} for the four samples. 
One can see that $H_{c}^{^\prime}(T)$ increases rapidly with decreasing $T$ and displays a linear 
variation near $T_{c}^{\prime}$. We have tentatively tried to estimate a low $T$ extrapolation of $H_{c}^{^\prime}$
using a parabolic $T$ variation as applied for the critical field of classical superconductors:
\begin{equation}
H_{c}^{\prime }(T)=H_{c}^{\prime }(0)[1-(T/T^{\prime }_{c})^{2}].
\label{Eq.H'c-T}
\end{equation}
The fitting curves are displayed as dashed lines in Fig.\ref{Fig:Hcprime} and show that
$H_{c}^{\prime}(0)$ increases with hole doping and reaches a value as high as $\sim 150$ Tesla 
at optimal doping.

A precise analysis of the fluctuation magnetoconductivity $\Delta\sigma_H(T,H)$ 
is a valuable tool to extract different microscopic parameters of high-$T_{c}$ cuprates, such as 
the value of $H_{c2}(0)$ not directly accessible from experiments. $\Delta\sigma_H(T,H)$ can be 
written out as:
\begin{eqnarray}
\Delta\sigma_{H}(T,H) & = & \Delta\sigma(T,H)-\Delta\sigma_{n}(T,H)\, \nonumber \\
& = & \Delta\sigma_{SF}(T,H)-\Delta\sigma_{SF}(T,0).
\label{Eq.correct-magneto}
\end{eqnarray}  
It has been very often assumed that the second term of the first equation can be neglected as being 
only weakly dependent on magnetic fields. However, our study clearly shows that this is not the case 
(see for instance the data displayed in fig.\ref{fig:MR-OPT}). Thus our method provides here a correct 
determination of $\Delta\sigma_H(T,H)$. 

In the GL approach, the evolution of the fluctuation
magnetoconductivity with $H$ comes from the pair-breaking effect which leads to a $T_c$ suppression. 
In the case of interest, the major contribution results from the AL process, and more particularly 
from the interaction of the field with the carrier orbital (ALO) degrees of freedom. The detailed 
analysis and discussion of the fluctuation magnetoconductivity are reported in ref.\cite{FRA-PRB2011}. The important
point here is that the analysis of the fluctuation magnetoconductivity can give a direct determination 
of the coherence length, and then of $H_{c2}(0)$, as the same coherence length governs the fluctuating and the superconducting regimes. Thus a mirror field $H^{\star}(T)$ of the upper critical field $H_{c2}(T)$ can be determined above $T_c$ in the GL approach \cite{Larkin-Varlamov}.
Fig.\ref{Fig:fit-ALO} shows the data for the UD85 sample together with the fits using the ALO 
expression with $H_{c2}(0)=125(5)$T being the only adjustable parameter. We checked that, 
as predicted by the theory, the fits are valid as long as $H \lesssim H^{\star}(T)=\epsilon H_{c2}(0)$. 

Let us mention here that the analysis of the magnetoconductivity in terms of the ALO expression has to be 
restricted to the temperature range where the fluctuating conductivity can be also well described within 
the GL framework. For instance for the UD-85 sample, it is not possible to fit the $\Delta\sigma_H(H,T)$ 
curves with the same value of $H_{c2}$ for $T>99$K ($\epsilon > 0.16$) above which $\Delta\sigma_{SF}(T)$ 
starts to deviate significantly from the LD expression (see fig.\ref{Fig:LD}). This would give values 
of $H_{c2}$ correspondingly smaller as the temperature is higher. Moreover, in the case of the UD57 sample, 
the fits of the $\Delta\sigma_H(H,T)$ curves can only be performed for temperatures above the mean field 
temperature $T_{c0}$ using the value of $T_{c0}=72$K in the ALO expression. It can be seen in table \ref{tab:H} 
that the value of $H_{c2}$ extracted from the low-field part of the magnetoconductivity data matches very well 
that of $H_c^{\prime}(0)$ obtained in a completely different way. Let us emphasize here that the comparison can 
only be indicative as the use of Eq.\ref{Eq.H'c-T} is not granted. So the value of $H_c^{\prime}(0)$ could as well
be a lower bound of the field above which the SCFs are suppressed. However, the fact that both fields are 
comparable for all the doping contents investigated highlights the consistency of our data analysis. 

\begin{table}[ht]
\caption{Values of $H_{c2}(0)$ extracted from the fluctuation magnetoconductivity. They are comparable 
to the extrapolated values of the onset field of SCF $H_{c}^{\prime}(0)$.(from ref.\cite{FRA-PRB2011})}
\label{tab:H}
\begin{center}     
\begin{tabular}{lllll}
\br
Sample & UD57 & UD85 & OPT93.6 & OD92.5  \\
\mr
$H_{c2}(0)$(T) & 90(10) & 125(5) & 180(10) & 200(10) \\
$H_{c}^{\prime}(0)$(T) & 86(10) & 115(5) & 155(10) & 207(10) \\
\br
\end{tabular}
\end{center}
\end{table}
The important result here is to show that the superconducting gap which is directly 
related to $H_{c2}(0)$ increases smoothly with increasing hole doping from the underdoped to the 
overdoped regime, contrary to the pseudogap which decreases. This is a strong indication that the gap determined here can thus be assimilated to the "small" gap detected recently by different techniques, while the pseudogap would be rather connected with the "large" gap \cite{Hufner}.

One can point out that rather different values of $H_{c2}$ have been reported in ref.\cite{Ando-Sega} from the analysis of the $T$ dependence of the magnetoconductivity at 1T in untwinned YBCO crystals. Even if this study was performed on different single crystals, one can conjecture that the very few data points used in that case to fit the fluctuation magnetoconductivity in a temperature range where the validity of the GL approach was not really checked, might be the source of large errors in the determination of $H_{c2}$.

\section{Influence of disorder}
\label{sec:disorder}
It is now well admitted that the properties of cuprates are strongly dependent on disorder. 
We have studied for long the effect of the introduction of controlled disorder by electron irradiation 
and the way it affects the transport properties \cite{FRA-PRL2003}. In particular, we have shown that similar
upturns of the low-T resistivity are found for controlled disorder in YBCO and in some "pure" low-$T_c$
cuprates, which indicates the existence of intrinsic disorder in those families \cite{FRA-EPL-MIT}. 

We have also carried out magnetoresistance measurements in some OPT93.6 and UD57 samples irradiated by
electrons. When $T_c$ is decreased by disorder, we find that both $T_{c}^{\prime}$ 
and $H_{c}^{\prime}(0)$ are also affected. The reduction in $T_c^{\prime}$ nearly follows that in $T_c$ 
for the underdoped sample while it is slightly larger for the OPT sample. Consequently, when $T_c$ is
decreased by disorder, the relative range of SCFs with respect to the value of $T_c$ expands considerably.
For instance, in ref.\cite{FRA-PRB2011}, we still detect $T_{c}^{\prime} \sim 60$K in an UD57 irradiated sample with $T_c = 4.5$K. 

\begin{figure}[h]
\includegraphics[width=16pc]{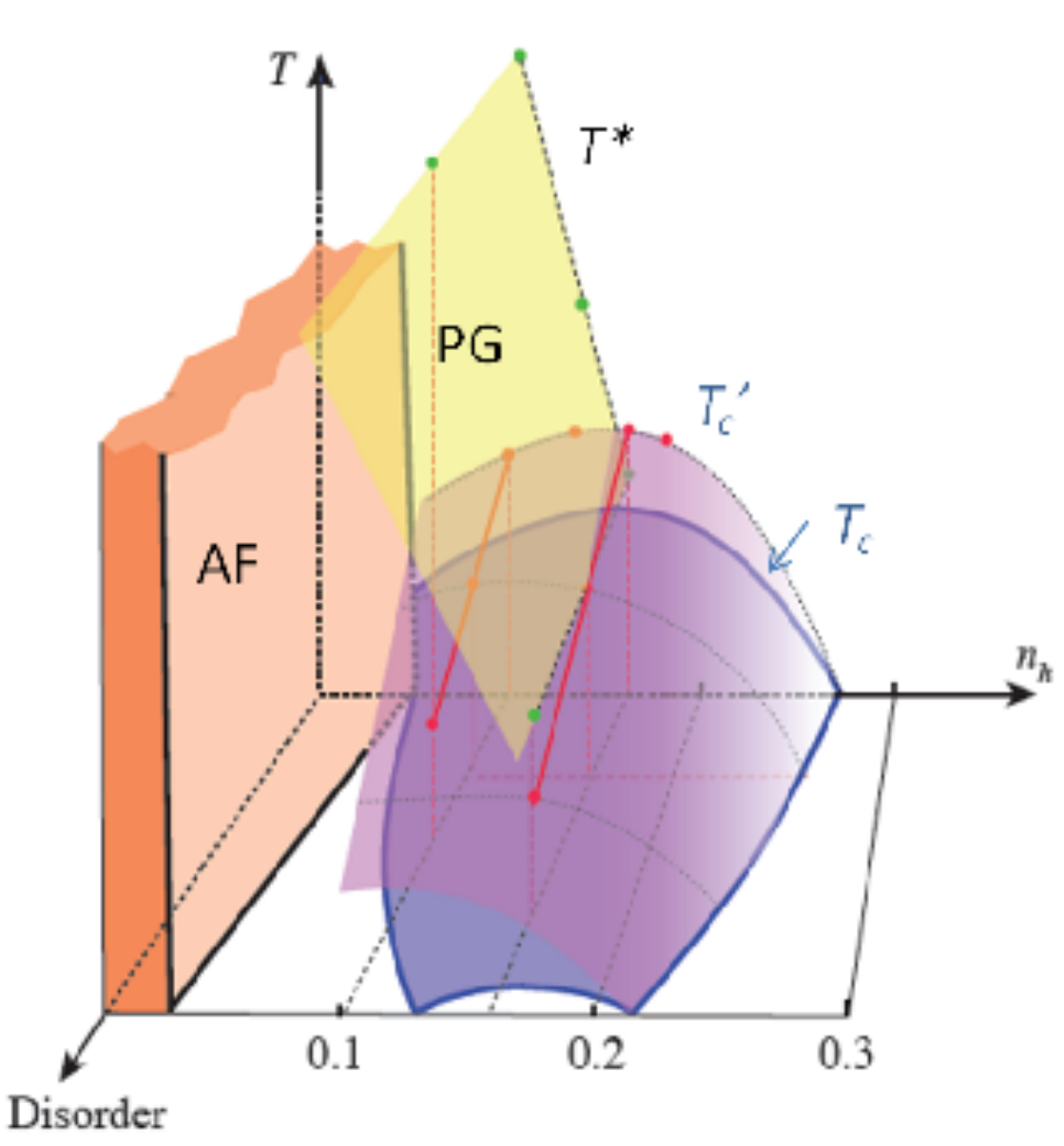}\hspace{2pc}%
\begin{minipage}[b]{16pc}\caption{\label{Fig:3Dphase-diagram}Phase diagram constructed on the data points
obtained here, showing the evolution of $T_{c}^{\prime}$ the onset of
SCF, with doping and disorder (from ref.\cite{FRA-PRB2011}). The fact that the pseudogap and the SCF
surfaces intersect each other near optimum doping in the clean limit is
apparent. These surfaces have been limited to experimental ranges where
they have been determined experimentally.}
\end{minipage}
\end{figure}

These results allow us to draw important conclusions on the cuprate phase diagram. Indeed, 
contrary to $T_c$, $T_c^{\prime}$ or $H_c^{\prime}$, the pseudogap temperature $T^{\star}$ has 
been found very early to be quite robust to disorder \cite{Alloul-PRL1991}. This is another indirect 
evidence that the pseudogap phase is not related to superconductivity. We want also to emphasize here 
that specific effects induced by disorder are probably at the origin of many confusions 
in the study of high-$T_c$ cuprates. This leads us to propose in fig.\ref{Fig:3Dphase-diagram} 
a 3D phase diagram where the effect of disorder has been introduced as a third axis. 

There, in the pure systems, the occurrence of SCFs and the difficulty to separate the SC gap from 
the pseudogap in zero-field experiments justifies that the $T_c^{\prime}$ line could often be taken as a 
continuation of the $T^{\star}$ line. 

It can also be seen in this figure that the respective evolutions with disorder 
of the SC dome and of the amplitude of the SCF range explains the phase diagram often shown in a 
low-$T_c$ cuprate such as Bi-2201 and sketched in fig.\ref{fig:phase-diagram}(a). Finally, for 
intermediate disorder, the enhanced fluctuation regime with respect to $T_c$ observed in the 
Nernst measurements for the La$_{2-x}$Sr$_x$CuO$_4$ can be reproduced as well \cite{Wang-PRB2006}.

\section{Conclusion}
\label{sec:conclusion}
We have presented here a condensed report of our quantitative study of the superconducting fluctuation conductivity in YBCO that was extensively detailed in previous publications 
\cite{FRA-HF}-\cite{Alloul-EPL2010}. Our data give important determinations of some 
thermodynamic properties of the SC state of 
high-$T_{c}$ cuprates which are not accessible otherwise, as flux flow dominates near $T_{c}$ in the 
vortex liquid phase and the highest fields available so far are not sufficient to overcome $H_{c2}$ 
and then to reach the normal state at $T=0$. 

The consistency of our data analyses, which establish that the SCFs do match quantitatively the expectations from the 2D GL approach, strongly justifies the method used to determine the superconducting fluctuation conductivity from the deviations of the magnetoresistance from the $H^2$ normal state observed well above $T_c$.In a metallic state, deviations of the MR from an $H^2$ behavior could of course be expected in case of Fermi surface reconstruction. A two-carrier model has been indeed proposed to explain some MR data in YBa$_2$Cu$_4$O$_8$ \cite{Rourke2} in which quantum oscillations could be detected at low $T$. In that approach the contributions of the SCFs has been completely ignored above $T_c$, though they should undoubtedly be present. 

For our three higher oxygen content samples around optimal doping, Fermi Surface reconstructions have never been observed so far. As for our underdoped sample, it has a lower doping and lower $T_c$ than the  ortho-II ordered YBCO$_{6.5}$ sample in which Fermi surface reconstruction has been detected at the highest $T$ of about 60K \cite{Doiron}. Our UD sample being twinned and thus with poor oxygen order, a  reconstruction, if any, should then occur at even lower $T$.  So, if such reconstruction effects do occur in this very sample, their influence should not extend in the high temperature range of our measurements (from 70K up to 130K). The  oxygen disorder could play a role in reducing the $T_c$ of that sample. As in our lower $T_c$ samples in which disorder has been introduced on purpose, this could justify that, in presence of disorder, we need to introduce a mean field $_Tc$ value higher than the zero field $T_c$ to interpret the data in a GL approach \cite{FRA-PRB2011}.

Finally, contrary to what was claimed by others in this conference \cite{M2S}, our data unambiguously show that the superconducting fluctuations vanish abruptly with increasing temperature, allowing us to define an onset temperature $T_c^{\prime}$ and an onset magnetic field $H_c^{\prime}$ above which the SCF contribution to conductivity becomes unmeasurable. The comparison between the huge drop of the electronic susceptibility determined by NMR from $T^{\star}$ and the emergence of superconducting fluctuations at $T_c^{\prime}$ well below $T^{\star}$ clearly indicates that these two temperature scales are not connected with each other. 

We therefore have evidenced that the fluctuation conductivity can be very well accounted for in terms of the first series expansion of gaussian fluctuations in a limited temperature range, but require extension of the theory to explain the sharp drop at higher $T$. Moreover, the analysis of the fluctuating magnetoconductivity in this temperature range allows us to demonstrate 
that the pairing energy and SC gap both increase 
with doping, confirming then that the pseudogap has to be assigned to an independent magnetic order 
or crossover due to the magnetic correlations.

\ack
We acknowledge collaboration with C. Proust, B. Vignolle and G. Rikken at the LNCMI.
This work has been performed within the "Triangle de la Physique" and was supported by 
ANR grants "OXYFONDA" NT05-4 41913 and "PNICTIDES". The experiments at LNCMI-Toulouse were funded by the FP7 I3 EuroMagNET.


\bigskip 
\begin{thebibliography}{20}
\bibitem{Alloul-1989}Alloul H Ohno T and Mendels P 1989 \textit{Phys. Rev. Lett.} \textbf{63}, 1700

\bibitem{Timusk}Timusk T and Statt B 1999 \textit{Rep. Prog. Phys.} \textbf{62} 61

\bibitem{Emery}Emery VJ and Kivelson SA 1995 \textit{Nature} \textbf{374} 434

\bibitem{Wang-PRB2006}Wang Y Li L and Ong NP 2006 \textit{Phys. Rev.} B \textbf{73} 024510

\bibitem{Hufner}H\"ufner S Hossain MA Damascelli A and Sawatzky GA 2008 \textit{Rep. Prog. Phys.} \textbf{71} 062501

\bibitem{FRA-HF}Rullier-Albenque F Alloul H Proust C Colson D and Forget A 2007 \textit{Phys. Rev. Lett.} \textbf{99} 027003

\bibitem{FRA-PRB2011}Rullier-Albenque F Alloul H Rikken G 2011 \textit{Phys. Rev.} B \textbf{84} 014522

\bibitem{Alloul-EPL2010}Alloul H Rullier-Albenque F Vignolle B Colson D and Forget A 2010 \textit{Europhys. Lett.} \textbf{91} 37005

\bibitem{Legris}Legris A Rullier-Albenque F.Radeva E and Lejay P 1993 \textit{J. Phys.} I 
France \textbf{3} 1605

\bibitem{Harris}Harris JM Yan YF Matl P Ong NP Anderson PW Kimura T and Kitazawa K 1995 \textit{Phys. Rev. Lett.} \textbf{75}, 1391 

\bibitem{FRA-Nernst}Rullier-Albenque F Tourbot R Alloul H Lejay P Colson D and Forget A 2006 \textit{Phys. Rev. Lett.} \textbf{96} 06700

\bibitem{Li} Li L Wang Y Komiya S Ono S Ando Y Gu GD and Ong NP 2010 \textit{Phys. Rev. B}\textbf{81} 054510

\bibitem{Grbic} Grbi\'c MS Po\v zek M Paar D Hinkov V Raichle M Haug D Keimer B Bari\v si\'c N and Dul\v ci\'c A 2011 \textit{Phys. Rev. B} \textit{83} 144508.

\bibitem{Bilbro} Bilbro LS Vald\'es Aguilar R Logvenov G Pelleg O Bo\v zovi\'c I and Armitage NP 2011 \textit{Nature Physics} \textbf{7} 298

\bibitem{Rourke} Rourke PMC Mouzopoulou I Xu X Panagopoulos C Wang Y Vignolle B Proust C Kurganova V Zeitler U Tanabe Y Adachi T Koike and Hussy NE 2011 \textit{Nature Physics} \textbf{7} 455

\bibitem{Wang-PRL2}Wang Y Li L  Naughton MJ Gu GD Uchida S and Ong NP 2005 \textit{Phys. Rev. Lett.} \textbf{95} 247002

\bibitem{Alloul-1993}Alloul H Mahajan A Casalta H Klein O (1993) \textit{Phys. Rev. Lett} \textbf{70} 1171

\bibitem{Larkin-Varlamov}For a review see Larkin A and Varlamov AA 2005 \textit{Theory of fluctuations in
superconductors} (Oxford University Press, Oxford.

\bibitem{Lawrence-Doniach}Lawrence WE and S. Doniach 1971 Proceedings 12th International Conference on Low
Temperature Physics, Kyoto 1970 (E. Kanda, Keigaku, Tokyo) 361

\bibitem{Reggiani} Reggiani l Vaglio R Varlamov A.A (1991) \textit{Phys. Rev. B} \textbf{44} 9541

\bibitem{Ando-Sega} Ando Y Segawa K (2002) \textit{Phys. Rev. Lett.} \textbf{88} 167005

\bibitem{FRA-PRL2003} Rullier-Albenque F Alloul H Tourbot R 2003 \textit{Phys. Rev. 
Lett.} \textbf{91} 047001

\bibitem{FRA-EPL-MIT} Rullier-Albenque F Alloul H Balakirev F Proust C. 2008 \textit{Europhys. 
Lett.} \textbf{81} 37008

\bibitem{Alloul-PRL1991} Alloul H Mendels P Casalta H Marucco JF Arabski J (1991) \textit{Phys. Rev. Lett.} \textbf{67} 3140

\bibitem{Rourke2}Rourke PMC. Bangura AF. Proust C Levallois J Doiron-Leyraud N LeBoeuf D Taillefer L Adachi S Sutherland SL and Hussey NE 2010 \textit{Phys. Rev. B} \textbf{82}, 020514(R)

\bibitem{Doiron}Doiron-Leyraud N Proust C LeBoeuf D Levallois J Bonnemaison JB Liang R Bonn DA Hardy WN Taillefer L 2007 \textit{Nature} \textbf{447} 565

\bibitem{M2S} Taillefer L, Chang J et al. talks at this M2S conference

\end{thebibliography}
\end{document}